\documentclass[12pt,aip,pop,preprint,footinbib]{revtex4-1}		
\usepackage{color}
\usepackage{graphicx}
\usepackage{float}
\usepackage{epsfig}
\usepackage{epstopdf}
\usepackage{subfig}
\usepackage{bm} 

\def\calL{{\cal L}}
\def\delStar{\Delta^*}
\def\grad{\nabla}

\def\div{\nabla\cdot}

\def\fsAver#1{\left\langle #1 \right\rangle} 

\def\begeqn{\begin{equation}}
\def\endeqn{\end{equation}}
\def\begeqnar{\begin{eqnarray}}
\def\endeqnar{\end{eqnarray}}
\def\begeqnarn{\begin{eqnarray*}}
\def\endeqnarn{\end{eqnarray*}}

\begin{document}
\title[]{Exact Solov'ev equilibrium with an arbitrary boundary} 
\author{A. Y. Aydemir}
\email{aydemir@nfri.re.kr}
\author{B. H. Park}     
\affiliation{National Fusion Research Institute, Daejeon 34133, Republic of Korea}   
\author{K. S. Han}
\affiliation{University of Science and Technology, Daejeon, Korea}                
\baselineskip 18pt  

\begin{abstract}
Exact Solov'ev equilibria for arbitrary plasma cross-sections are calculated using a constrained least-squares method. The boundary, with or without $X$-points,  can be specified with an arbitrarily large number of constraints to ensure an accurate representation. Thus, the order of the polynomial basis functions in the homogeneous solution of the Grad-Shafranov equation becomes an independent parameter determined only by the accuracy requirements of the overall solution. Examples of exact, highly-shaped equilibria are presented.
\end{abstract} 
\maketitle 

\section{Introduction}
An axisymmetric equilibrium, starting point for many linear stability and nonlinear initial-value calculations for toroidal devices, is described by the Grad-Shafranov equation \cite{grad1958, shafranov1958},
\begeqn
-\delStar\psi = \mu_0R^2p'(\psi) + FF'(\psi), \label{eqn:GS0}
\endeqn
where primes denote a derivative with respect to $\psi$. In cylindrical $(R,Z,\zeta)$ coordinates, where the toroidal angle $\zeta$ is in the symmetry direction, the Laplacian-like operator $\delStar$ has the form
\begeqn
\delStar\psi \equiv R^2\div\frac{1}{R^2}\grad\psi = R\frac{\partial}{\partial R}\frac{1}{R}\frac{\partial\psi}{\partial R} + \frac{\partial^2\psi}{\partial Z^2}. \label{eqn:delStar}
\endeqn
The equilibrium is determined by specifying the two source functions on the right, $p(\psi), F(\psi)$, for the plasma pressure and poloidal current, respectively.

Although the original references are not easily accessible anymore, derivation of Eq.~\ref{eqn:GS0} can be found in many textbooks (see, for example Jardin\cite{jardin2010}, Freidberg\cite{freidberg2014}). The review article by Takeda and Tokuda\cite{takeda1991}, although somewhat dated, is also still useful. In ``fixed-boundary'' equilibria,  a boundary curve $C_b$ in the $(R,Z)$ (poloidal) plane is also specified a priori. Then Eq.~\ref{eqn:GS0} is solved with a Dirichlet boundary condition on $C_b$, typically $\psi=0.$ In ``free-boundary'' equilibria, a distribution of external currents in the poloidal field (PF) coils is given and the plasma boundary is determined as part of the solution. Here, effectively a boundary condition at infinity needs to be imposed, $\psi\rightarrow 0$. Usually this problem in the infinite domain is tackled using an artificial computational boundary, $C_{cb}$,  a finite distance away from the plasma. Recently we have introduced a new method that directly and efficiently solves the free-boundary problem in the infinite domain\cite{hanks2019b} without a need for $C_{cb}$. Note that besides $p(\psi)$ and $F(\psi)$, a more general pair of source functions can also be used as input\cite{lutjens1996}. In fact, widely-used EFIT-series of equilibrium codes\cite{lao2005} can calculate equilibria consistent with a wide range of experimental data from kinetic and magnetic diagnostics.

A realistic equilibrium that satisfies a number of constraints, for example, for the plasma pressure and safety factor profiles, generally requires $p$ and $F$ to be nonlinear functions of their argument. Therefore, Eq.~\ref{eqn:GS0} is typically a nonlinear equation for the flux function $\psi$, and it is solved numerically. Non-numerical solutions, however, are quite useful in analytic equilibrium and stability studies; they are also very useful in checking and benchmarking of numerical codes. For these reasons, exact analytic solutions of the Grad-Shafranov equation have been sought by a number of workers over the years\cite{mcCarthy1999, atanasiu2004, guazzotto2007}. The most well-known and useful of these, however, is a class of exact solutions that have come to be known as the Solov'ev equilibria\cite{solovev1967}, which assume $p$ and $F^2$ are linear functions of $\psi$. In Eq.~\ref{eqn:GS0} letting $\mu_0p'=-C,~FF'=-AR_0^2,$ where
$A,C$ are constants and $R_0$ is the major radius of the torus, leads to the inhomogenous but linear equation 
\begeqn
\delStar\psi = AR_0^2 + CR^2, \label{eqn:GS1}
\endeqn
with the exact Solov'ev solution
\begeqn
\psi = \frac{1}{2}(AR_0^2 + a_0R^2)Z^2 + \frac{1}{8}(C-a_0)(R^2-R_0^2)^2, \label{eqn:gs1Soln}
\endeqn
where $a_0$ is another constant. The three arbitrary constants, $a_0, A, C$, can be adjusted to change, for instance, the separatrix geometry\cite{solovev1967}. However,  there are not enough degrees of freedom in the solution to specify basic equilibrium parameters such as the plasma $\beta=2\mu_0p/B^2$, safety factor $q$, and at the same time ensure that the solution conforms to a particular plasma boundary.

Solutions of the homogeneous equation, 
\begeqn
\delStar\psi=0, \label{eqn:homogeneous}
\endeqn
can provide the needed extra degrees of freedom, since a general solution to Eq.~\ref{eqn:GS1} can be written as
$\psi = \psi_p + \psi_h$, where $\psi_p$ is a particular solution of  Eq.~\ref{eqn:GS1}, and $\psi_h$ satisfies Eq.~\ref{eqn:homogeneous}.
A complete set of multipole solutions of Eq.~\ref{eqn:homogeneous} were found by Reusch and Nielson and applied to PDX vacuum field calculations\cite{reusch1986}. A similar polynomial expansion in powers of  $R,Z$ was used by Zheng {\em et al.}\cite{zhengSB1996} to find shaped but up-down symmetric, exact analytic solutions to Eq.~\ref{eqn:GS1}. These were later generalized by Cerfon and Freidberg\cite{cerfon2010} to geometries without up-down symetry, including diverted configurations with one or more $X$-points.

In both Zheng\cite{zhengSB1996} and Cerfon\cite{cerfon2010}, the number of constraints, $M$, used to specify the boundary was limited by the degrees of freedom, $N$,  in the homogeneous solution. Thus, a more detailed boundary definition required a larger $N$, which translated into a higher order polynomial expansion for the homogeneous solution $\psi_h.$ In this work we show how the requirements of the boundary definition can be decoupled from the  polynomial expansion of $\psi_h$ using constrained least-squares. Now the boundary, both ``fixed" and ``free,'' can be defined essentially arbitrarily while still maintaining a fixed and small number of terms in the expansion of $\psi_h.$ The method is explained in Sect. II with various examples of exact tokamak equilibria. Sect. III summarizes our results.

\section{Exact solutions with an arbitrary boundary}
It is useful to put Eq.~\ref{eqn:GS1} in nondimensional form by transforming the variables. After letting $\psi = (B_0R_0^2){\widetilde\psi}$, $R = R_0{\widetilde R}$, $Z=R_0{\widetilde Z}$, the equation becomes
\begeqn
\delStar\psi = A + CR^2, \label{eqn:GS1norm}
\endeqn
where we dropped the tildes in normalized variables. Typically $B_0$ is the toroidal field strength at the major radius $R_0.$ A $Z$-independent particular solution is given by
\begeqn
\psi_p = \frac{C}{8}R^4 + \frac{A}{2}R^2\ln{R}. \label{eqn:psiP}
\endeqn
The constants $A,C$ can be used to specify, for example,  the plasma $\beta$ and the kink safety factor $q^*=\epsilon B_0/B_p,$ where $\epsilon\equiv a/R_0$ is the inverse aspect ratio, and $B_p$ is the poloidal field at the edge. As stated earlier,  to satisfy the constraints that help define the boundary, we will need the degrees of freedom provided by the homogeneous solution.

\subsection{Homogeneous solution}
Following Zheng {\em et al.}\cite{zhengSB1996}  we write the homogeneous solution in the form
\begeqn
\psi_h = \sum_{i=0} f_i(R)Z^i. \label{eqn:expansion}
\endeqn
Substituting in Eq.~\ref{eqn:homogeneous} leads to the recursion relation
\begeqn
R\frac{\partial}{\partial R}\frac{1}{R}\frac{\partial}{\partial R}f_{i-2}(R) =-i(i-1)f_i(R),~i \ge 2.
\endeqn
Clearly the even and odd terms decouple. Letting $f_i(R) = 0$ for $i\ge I$ and integrating yields a solution with $N=2I$ constants of integration (we will assume $I$ is even). Then the homogeneous solution can be written as a sum of polynomial basis functions:
\begeqn
\psi_h(R,Z) = \sum_{i=0}^{I-1}c_iP_i(R,Z) + \sum_{i=1}^{I}d_iQ_i(R,Z),  \label{eqn:expansion2}
\endeqn
where $P_i$ and  $Q_i$ have even and odd order, respectively. For up-down symmetric boundaries,  the second sum is not needed. The resulting polynomials for $I=10$, obtained using {\em Mathematica}\cite{mathematica2019}, are shown in the Appendix. The set of arbitrary coefficients, 
\begeqn
\{{c_0,c_1,\ldots, c_{I-1}, d_1,d_2,\ldots,d_I}\},
\endeqn
will be determined by the boundary constraints to be explained in the next section. Since the coefficients are arbitrary, each polynomial basis function has to satisfy the homogeneous equation identically, which can be shown to be true by direct substitution.
Below we first consider ``limited'' or fixed-boundary tokamak equilibria without $X$-points.

\subsection{Exact fixed-boundary equilibria}
\begin{figure}[htbp]
\begin{center}
\includegraphics[width=2in]{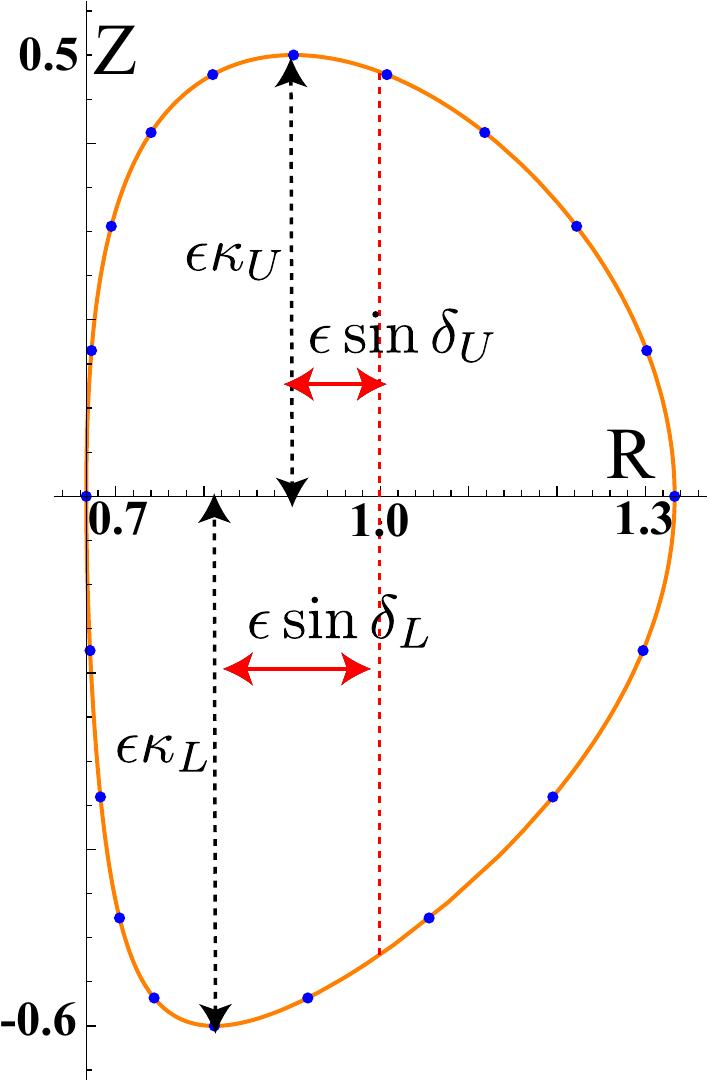}
\caption{\em \baselineskip 14pt The boundary curve for $\epsilon=a/R_0=1/3, \delta_U=0.3,~\delta_L=0.6,~\kappa_U=1.5,~\kappa_L=1.8.$}
\label{fig:geometry}
\end{center}
\end{figure}

A common parametrization of shaped plasma boundaries has the form
\begeqnar
R(\alpha) & = & 1+ \epsilon\cos[\alpha + \delta(\alpha)\sin\alpha], \nonumber \\
Z(\alpha) & = & \epsilon\kappa(\alpha)\sin\alpha, \label{eqn:parametricCurve}
\endeqnar
where $(R,Z)$ have been normalized by $R_0$. The triangularity $\delta$ and elongation $\kappa$ are defined in terms of their upper and lower values as:
\begeqnar
\delta(\alpha) & = & \frac{1}{2}\left[(\delta_U + \delta_L) + (\delta_U -\delta_L)\sin\alpha\right], \nonumber \\
\kappa(\alpha) & = & \frac{1}{2}\left[(\kappa_U + \kappa_L) + (\kappa_U -\kappa_L)\sin\alpha\right]. \label{eqn:alpha-kappa}
\endeqnar
The geometric meanings of various terms are shown in Fig.~\ref{fig:geometry}. Note that the parameter $\alpha$ differs from the angle $\theta=\arctan(Z/(R-R_0)).$ The actual triangularities, which measure the shift of the highest and lowest points of the boundary with respect to the geometric center, are $\Delta_U=\sin\delta_U\simeq \delta_U$, and $\Delta_L=\sin\delta_L\simeq\delta_L.$

A useful set of boundary constraints uses an equally spaced points on the boundary curve, as shown in Fig.~\ref{fig:geometry}, and take the following form:
\begeqnar
\psi(R_j,Z_j) & = & \psi_p(R_j,Z_j) + \psi_h(R_j,Z_j) =  \psi_b, \label{eqn:boundaryConst} \\
(R_j, Z_j)  & \equiv & (R(\alpha_j),Z(\alpha_j)),~\alpha_j = \frac{2\pi j}{M},~j=0,1,\ldots M-1, \nonumber
\endeqnar
where $\psi_b$ is the boundary value; we will set $\psi_b=0.$

If the number of constraints exactly matches the degrees of freedom in the homogeneous solution, i.e., if $M=N$, then the set of equations in Eq.~\ref{eqn:boundaryConst} can be solved easily for the $N=2I$ coefficients ($c_i,d_i)$ of Eq.~\ref{eqn:expansion2}, a method adopted by the earlier works\cite{zhengSB1996, cerfon2010}. However, there are obvious advantageous to having $M \gg N$. A highly-shaped boundary may require many more points to specify accurately than the degrees of freedom available in the homogeneous solution. But with $M>N$, the system is overdetermined and the linear system resulting from Eq.~\ref{eqn:boundaryConst} has no solution.

\subsection{Least squares solution}
We can still find a solution for $M>N$ if we do not demand that all points be located exactly on the boundary, i.e., we stop requiring that Eq.~\ref{eqn:boundaryConst} be satisfied exactly.  Instead we seek the ``best'' solution that minimizes, for example, a positive-definite error term. In the least-squares method we minimize the sum of the ``residuals,'' defined by
\begeqnar
S & = & \frac{1}{2}\sum_{j=0}^{M-1}w_j\left\{\psi(R_j,Z_j)-\psi_b\right\}^2 \nonumber \\
& = & \frac{1}{2}\sum_{j=0}^{M-1}w_j\left\{\sum_{i=0}^{I-1}c_iP_i(R_j,Z_j) + \sum_{i=1}^{I}d_iQ_i(R_j,Z_j) + \psi_P(R_j,Z_j)- \psi_b \right\}^2 , \label{eqn:Sum}
\endeqnar
where $(R_j,Z_j),~j=0,1,\ldots, M-1$ are points on the boundary parametrically determined by Eq.~\ref{eqn:parametricCurve}, and $w_j \equiv w(R_j,Z_j)$ is a weight function that can be used to emphasize/de-emphasize some subset of the boundary points. It is not needed for fixed-boundary equilibria and is set to unity. 

Letting $\psi_b=0$ and seeking a minimum by setting $\partial S/\partial c_k=0,~\partial S/\partial d_l=0$ lead to
\begeqnar
\sum_{i=0}^{I-1}c_i\fsAver{P_i,P_k} + \sum_{i=1}^{I}d_i\fsAver{Q_i,P_k} & = & - \fsAver{\psi_P,P_k},~k=0,1,\ldots, I-1,\label{eqn:linearSetP} \\
\sum_{i=0}^{I-1}c_i\fsAver{P_i,Q_l} + \sum_{i=1}^{I}d_i\fsAver{Q_i,Q_l} & = & - \fsAver{\psi_P,Q_l},~l=1,2,\ldots, I, \label{eqn:linearSetQ}
\endeqnar
where we defined
\begeqn
\fsAver{p,q} \equiv \sum_{j=0}^{M-1}w(R_j,Z_j)p(R_j,Z_j)q(R_j,Z_j).
\endeqn
Together Eqs.~\ref{eqn:linearSetP}, \ref{eqn:linearSetQ} form a set of $N=2I$ linear equations for the $N$ coefficients $c_i,d_i$ that can be easily inverted. Again, if the boundary curve is symmetric about the mid-plane, we can set $d_i=0$ above and solve only for the $N=I$ coefficients $c_i$. In this case the parameter $\alpha_j$ in Eq.~\ref{eqn:boundaryConst} can be redefined to cover, for instance, only the lower half of the boundary.

\begin{figure}[htbp]
\begin{center}
\includegraphics[width=6in]{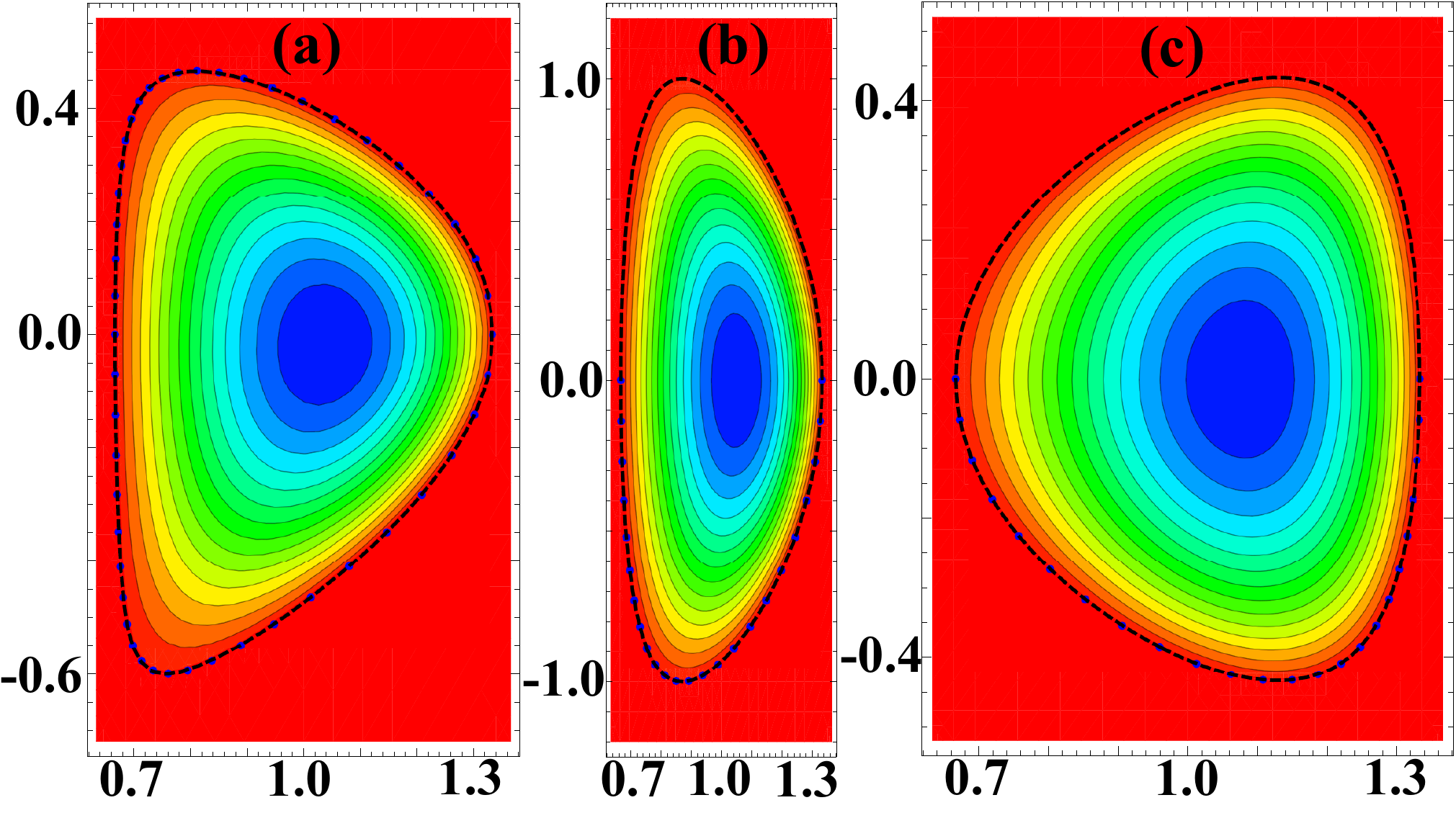}
\caption{\em \baselineskip 14pt Various highly-shaped equilibria. The (blue) dots are on the specified boundary curve (Eq.~\ref{eqn:parametricCurve}). The dashed (black) curve is the actual calculated boundary. (a) $\delta_U=0.6$, $\delta_L=0.8$, $\kappa_U=1.4$, $\kappa_L=1.8$. Calculated using $N=2I=12$, degrees of freedom in the polynomial expansion in Eq.~\ref{eqn:expansion2}, and $M=48$, the number of points (constraints) on the boundary (Eq.~\ref{eqn:boundaryConst}). (b) $\delta=0.4$, $\kappa=3, N=I=6$, $M=24$. (c) An equilibrium with negative triangularity: $\delta=-0.4$, $\kappa=1.3, N=I=6, M=24$. For all three equilibria, $\epsilon=a/R_0=1/3$, and the particular solution (Eq.~\ref{eqn:psiP}) has $A=0, C=1$.}
\label{fig:limCombined}
\end{center}
\end{figure}

Examples of highly-shaped exact equilibria obtained with this method are shown in Fig.~\ref{fig:limCombined}. For each one, $M$ number of boundary constraints are specified at the indicated points (blue dots in the figures) using Eq.~\ref{eqn:boundaryConst} and the parametric curve of Eq.~\ref{eqn:parametricCurve}.  For the asymmetric case in (a), both even and odd basis functions are used in the expansion of $\psi_h$; thus, there are  $N=2I$ degrees of freedom in the solution, where $I=6$ is the degree of the highest-order even basis function in Eq.~\ref{eqn:expansion2}.  For the up-down symmetric cases ((b) and (c)) only points in the lower half-plane are used. For these, the odd basis functions are not needed; thus they both have $N=I=6$. For all equilibria we use $M=4N$, so the resulting linear system of equations for the coefficients would be overdetermined without the least squares approach. In the figures, the actual calculated boundary is shown with a dashed (black) curve. Clearly the specified boundary (dots) and the actual boundary curve agree very well, although the least squares solution to the boundary constraints is not exact but only minimizes the sum of residuals in Eq.~\ref{eqn:Sum}. Note that, although not shown here, all solutions with $I=4$ (maximum order of even polynomials) also produce acceptable results for these highly-shaped fixed-boundary equilibria. 

\subsection{Exact equilibria with $X$-points}
An $X$-point, where the poloidal field vanishes, introduces three additional constraints that the solution $\psi=\psi_p + \psi_h$ has to satisfy:
\begeqnar
\psi(R_X,Z_X) & = & \psi_b, \label{eqn:XpointEqs} \\
\left.\frac{\partial \psi}{\partial R}\right|_{(R_X,Z_X)} & = & 0,~~~\left.\frac{\partial \psi}{\partial Z}\right|_{(R_X,Z_X)}  =  0, \nonumber
\endeqnar
where  $(R_X,Z_X) $ is the $X$-point location. Now  the separatrix becomes the plasma boundary; for simplicity we still choose $\psi_b\equiv \psi_{sep}=0.$

These three constraints, however,  are of a different nature than the boundary constraints in Eq.~\ref{eqn:boundaryConst}; they have to be satisfied exactly (otherwise there is no $X$-point), whereas those in Eq.~\ref{eqn:boundaryConst} can be satisfied in a ``least-squares'' sense--we can tolerate the computational boundary agreeing with the specified boundary only approximately to accommodate the $X$-point(s). Thus, we now have a  ``constrained least-squares'' problem, which we solve by introducing a ``Lagrangian $\calL$'' and the Lagrange multipliers $\lambda_j,~j=1,2,3$:
\begeqn
\calL({\bm c}, {\bm d}, {\bm \lambda}) = S + \lambda_1\left.\frac{\partial \psi}{\partial R}\right|_{(R_X,Z_X)} + \lambda_2\left.\frac{\partial \psi}{\partial Z}\right|_{(R_X,Z_X)} +\lambda_3\psi(R_X,Z_X), \label{eqn:lagrangian}
\endeqn
where the sum of residuals $S$ is still given by Eq.~\ref{eqn:Sum}. The unknown coefficient vectors $\bm c = \{c_0, c_1,\ldots, c_{I-1}\}$,  $\bm d = \{d_1, d_2,\ldots, d_{I}\}$ and $\bm \lambda = \{\lambda_1, \lambda_2, \lambda_3\}$ form the new set of $N+3$ unknowns.
They are determined by an augmented set of equations:
\begeqnar
\frac{\partial\calL}{\partial c_k} & = & \frac{\partial S}{\partial c_k} + \lambda_1\frac{\partial}{\partial c_k}\left(\frac{\partial \psi}{\partial R}\right)_{(R_X,Z_X)} + \lambda_2\frac{\partial}{\partial c_k}\left(\frac{\partial \psi}{\partial Z}\right)_{(R_X,Z_X)} + \lambda_3\frac{\partial \psi(R_X,Z_X)}{\partial c_k} = 0, \nonumber \\
& &  k=0,1,\ldots, I-1, \label{eqn:cEqs} \\
\frac{\partial\calL}{\partial d_l} & = & \frac{\partial S}{\partial d_l} + \lambda_1\frac{\partial}{\partial d_l}\left(\frac{\partial \psi}{\partial R}\right)_{(R_X,Z_X)} + \lambda_2\frac{\partial}{\partial d_l}\left(\frac{\partial \psi}{\partial Z}\right)_{(R_X,Z_X)} + \lambda_3\frac{\partial \psi(R_X,Z_X)}{\partial d_l} = 0, \nonumber \\
& &  l=1,2,\ldots, I, \label{eqn:dEqs} \\
\frac{\partial\calL}{\partial\lambda_1} & = & \left(\frac{\partial \psi}{\partial R}\right)_{(R_X,Z_X)} =0, \label{eqn:lambda1}\\
\frac{\partial\calL}{\partial\lambda_2} & = & \left(\frac{\partial \psi}{\partial Z}\right)_{(R_X,Z_X)} =0, \label{eqn:lambda2}\\
\frac{\partial\calL}{\partial\lambda_3} & = & \psi(R_X,Z_X) = 0. \label{eqn:lambda3}
\endeqnar
These equations guarantee that, among the set of solutions $\{{\bm c}, {\bm d}\}$ that satisfy the $X$-point constraints in Eqs.~\ref{eqn:lambda1}-\ref{eqn:lambda3} exactly, we choose the one that minimizes the residual sum $S$. Recall that $N=2I$ for up-down asymmetric equilibria. For symmetric ones, the coefficients $\bm d$ for the odd basis functions and Eq.~\ref{eqn:dEqs} are not needed, and $N=I$ (see Eq.~\ref{eqn:expansion2}).

We find that de-emphasizing the points in the sum of residuals $S$ near an $X$-point  leads to better results. Thus, if there is a field-null, we choose a Gaussian weight function in Eq.~\ref{eqn:Sum}  of the form
\begeqn
w(R_j,Z_j) = 1 - e^{-\{[(R_j-R_X)^2+(Z_j-Z_X)^2]/\Delta w^2\}},
\endeqn
where $\Delta w$ is an appropriately-chosen width and $(R_X,Z_X)$ is the location of the null-point. Note that $w\rightarrow 0$ near the $X$-point so that those points are less-constrained to be on the parametrically-defined boundary curve of Eq.~\ref{eqn:parametricCurve}, which is necessary to accommodate the $X$-point.

\begin{figure}[htbp]
\begin{center}
\includegraphics[width=6in]{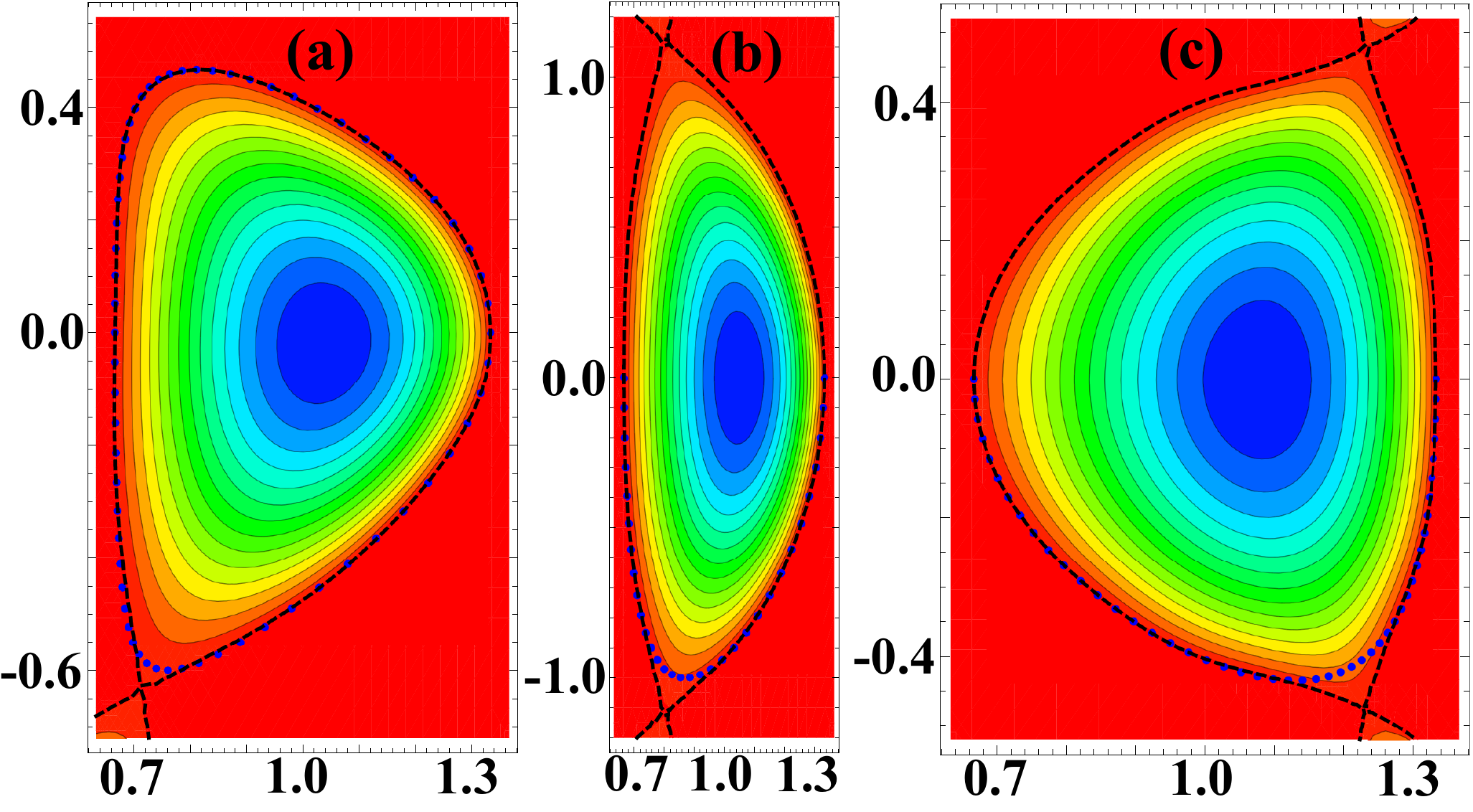}
\caption{\em \baselineskip 14pt Exact ``free-boundary'' versions of the fixed-boundary equilibria shown in Fig.~\ref{fig:limCombined}. The (blue) dots are on the specified boundary curve (Eq.~\ref{eqn:parametricCurve}). The dashed (black) curve is the actual calculated boundary. (a) $\delta_U=0.6$, $\delta_L=0.8$, $\kappa_U=1.4$, $\kappa_L=1.8$. Calculated using $N=2I=16$, degrees of freedom in the polynomial expansion in Eq.~\ref{eqn:expansion2}, and $M=64$, the number of points (constraints) on the boundary (Eq.~\ref{eqn:boundaryConst}). (b) $\delta=0.4$, $\kappa=3, N=I=8$, $M=32$. (c) An equilibrium with negative triangularity: $\delta=-0.4$, $\kappa=1.3, N=I=12, M=48$. For all three equilibria, $\epsilon=a/R_0=1/3$, and the particular solution (Eq.~\ref{eqn:psiP}) has $A=0, C=1$.}
\label{fig:XCombined}
\end{center}
\end{figure}

Examples of exact equilibria with $X$-points--``free-boundary'' versions of the fixed-boundary equilibria in Fig.~\ref{fig:limCombined}--are shown in Fig.~\ref{fig:XCombined}. We note that in all three cases, the computed boundary (dashed black curve) conforms to the specified boundary (represented by blue dots) very closely, except near the $X$-points where it necessarily deviates in order to satisfy the $X$-point constraints. Thus, even with one or more $X$-points, an exact equilibrium with a parametrically-specified boundary can be calculated using constrained least-squares method. However, all three examples require higher-order polynomial expansions than their fixed-boundary counterparts in Fig.~\ref{fig:limCombined}. The lower single null (LSN) configuration in (a) used an expansion with 16 degrees of freedom, or equivalently, the highest-order even basis function had order $I=8.$ Similarly, $I=8$ for (b), the elongated double-null (DN). The DN configuration with the negative triangularity in (c) required $I=12.$

\section{Summary and discussion}
We presented a method to calculate exact Solov'ev equilibria with or without $X$-points for arbitrary plasma cross-sections. Using a least-squares approach to enforce the boundary constraints essentially decouples the accuracy requirements of the boundary specification from the order of the polynomial expansion of the homogeneous solution for the Grad-Shafranov equation. Thus, a much larger number of boundary constraints can be used than the number of degrees of freedom available in the homogeneous solution, ensuring that the numerically-obtained boundary conforms to the specified boundary with high fidelity. Usefulness of the method was illustrated with examples of fixed-boundary equilibria with highly-shaped geometries and their generalization to configurations with one or more $X$-points. 

These exact solutions, which require nothing more than the inversion of a relatively small linear system of equations, can be easily performed with a symbolic/numerical algebra system such as {\em Mathematica} or {\em Matlab.} They should be highly useful, for example, in benchmarking new equilibrium codes. They may also prove useful in semi-analytic equilibrium and stability calculations.

\section*{Acknowledgements}
This work was supported by MSIP, the Korean Ministry of Science, ICT and Future Planning, through the KSTAR project.
Most of the symbolic and numerical calculations were performed with {\em Mathematica.}

\section*{Appendix}
The polynomials $P_i(R,Z), Q_i(R,Z)$ of Eq.~\ref{eqn:expansion2} for $I=10$, obtained with  {\em Mathematica}\cite{mathematica2019}:
\begeqnar
P_0 & : & 1 \\
P_1 & : & \frac{R^2}{2}\nonumber \\
P_2 & : & \frac{R^2}{2}+R^2 (-\log (R))+Z^2\nonumber \\
P_3 & : & \frac{R^2 Z^2}{2}-\frac{R^4}{8}\nonumber \\
P_4 & : & -\frac{15 R^4}{8}+\frac{3}{2} R^4 \log (R)+3 R^2 Z^2-6 R^2 Z^2 \log (R)+Z^4\nonumber \\
P_5 & : & \frac{R^6}{16}-\frac{3 R^4 Z^2}{4}+\frac{R^2 Z^4}{2}\nonumber \\
P_6 & : & \frac{25 R^6}{8}-\frac{15}{8} R^6 \log (R)-\frac{225 R^4 Z^2}{8}+\frac{45}{2} R^4 Z^2 \log (R)+\frac{15 R^2 Z^4}{2}-15 R^2 Z^4 \log (R)+Z^6\nonumber \\
P_7 & : & -\frac{5 R^8}{128}+\frac{15 R^6 Z^2}{16}-\frac{15 R^4 Z^4}{8}+\frac{R^2 Z^6}{2}\nonumber \\
P_8 & : & -\frac{1645 R^8}{384}+\frac{35}{16} R^8 \log (R)+\frac{175 R^6 Z^2}{2}-\frac{105}{2} R^6 Z^2 \log (R) \nonumber \\
& & -\frac{525 R^4 Z^4}{4}+105 R^4 Z^4 \log (R)+14 R^2 Z^6-28 R^2 Z^6 \log (R)+Z^8\nonumber \\
P_9 & : & \frac{7 R^{10}}{256}-\frac{35 R^8 Z^2}{32}+\frac{35 R^6 Z^4}{8}-\frac{7 R^4 Z^6}{2}+\frac{R^2 Z^8}{2} \nonumber \\
Q_1 & : & Z \\
Q_2 & : & \frac{R^2 Z}{2}\nonumber \\
Q_3 & : & \frac{3 R^2 Z}{2}-3 R^2 Z \log (R)+Z^3\nonumber \\
Q_4 & : & \frac{R^2 Z^3}{2}-\frac{3 R^4 Z}{8}\nonumber \\
Q_5 & : & -\frac{75 R^4 Z}{8}+\frac{15}{2} R^4 Z \log (R)+5 R^2 Z^3-10 R^2 Z^3 \log (R)+Z^5\nonumber \\
Q_6 & : & \frac{5 R^6 Z}{16}-\frac{5 R^4 Z^3}{4}+\frac{R^2 Z^5}{2}\nonumber \\
Q_7 & : & \frac{175 R^6 Z}{8}-\frac{105}{8} R^6 Z \log (R)-\frac{525 R^4 Z^3}{8}+\frac{105}{2} R^4 Z^3 \log (R)+\frac{21 R^2 Z^5}{2}-21 R^2 Z^5 \log (R)+Z^7\nonumber \\
Q_8 & : & -\frac{35 R^8 Z}{128}+\frac{35 R^6 Z^3}{16}-\frac{21 R^4 Z^5}{8}+\frac{R^2 Z^7}{2}\nonumber \\
Q_9 & : & -\frac{4935 R^8 Z}{128}+\frac{315}{16} R^8 Z \log (R)+\frac{525 R^6 Z^3}{2}-\frac{315}{2} R^6 Z^3 \log (R) \nonumber \\
& & -\frac{945 R^4 Z^5}{4}+189 R^4 Z^5 \log (R)+18 R^2 Z^7-36 R^2 Z^7 \log (R)+Z^9\nonumber \\
Q_{10} & : & \frac{63 R^{10} Z}{256}-\frac{105 R^8 Z^3}{32}+\frac{63 R^6 Z^5}{8}-\frac{9 R^4 Z^7}{2}+\frac{R^2 Z^9}{2}. \nonumber
\endeqnar
No attempt has been made to simplify or normalize the polynomials. They are presented in the form found by {\em Mathematica.} Note that  $P_{I-1}$ has order $I$, while $Q_I$ has order $I+1$.


\end{document}